\documentclass[aps,prl,reprint,preprintnumbers,superscriptaddress,amsmath,amssymb,bibnotes,longbibliography]{revtex4-2}
\usepackage{graphicx}
\usepackage{dcolumn}
\usepackage{bm}
\usepackage{color}
\usepackage{booktabs}
\usepackage{siunitx} 
\usepackage[colorlinks,linkcolor=blue,anchorcolor=blue,citecolor=blue,breaklinks,CJKbookmarks=True,urlcolor=blue,filecolor=blue,menucolor=blue,runcolor=blue]{hyperref}
\usepackage[mathlines]{lineno}

\begin{document}

\title{Hidden magnetic order within the pressure induced superconducting dome of UTe$_2$}

\author{Kaixin Ye}
 \affiliation{New Cornerstone Science Laboratory, Center for Correlated Matter
 	and School of Physics, Zhejiang University, Hangzhou 310058, China}%

\author{Lubin Wang}
\affiliation{New Cornerstone Science Laboratory, Center for Correlated Matter
	and School of Physics, Zhejiang University, Hangzhou 310058, China}%

\author{Dengpeng Yuan}
\affiliation{National Key Laboratory of Surface Physics and Chemistry, Mianyang 621908, China}%

\author{Binbin Zhang}
\affiliation{Nanhu Laser Laboratory, Changsha 410073, China}%

\author{Yanan Zhang}
\affiliation{New Cornerstone Science Laboratory, Center for Correlated Matter
	and School of Physics, Zhejiang University, Hangzhou 310058, China}%

\author{Ye Chen}
\affiliation{New Cornerstone Science Laboratory, Center for Correlated Matter
	and School of Physics, Zhejiang University, Hangzhou 310058, China}%

\author{Yu Liu}
\affiliation {New Cornerstone Science Laboratory, Center for Correlated Matter
	and School of Physics, Zhejiang University, Hangzhou 310058, China}

\author{Xin Lu}
\affiliation{New Cornerstone Science Laboratory, Center for Correlated Matter
	and School of Physics, Zhejiang University, Hangzhou 310058, China}

\author{Chaofan Zhang}
\affiliation{Nanhu Laser Laboratory, Changsha 410073, China}%

\author{Qiuyun Chen}
\affiliation{National Key Laboratory of Surface Physics and Chemistry, Mianyang 621908, China}

\author{Shiyong Tan}
\affiliation{National Key Laboratory of Surface Physics and Chemistry, Mianyang 621908, China}

\author{Frank Steglich}
\affiliation{New Cornerstone Science Laboratory, Center for Correlated Matter
	and School of Physics, Zhejiang University, Hangzhou 310058, China}
\affiliation{Max Planck Institute for Chemical Physics of Solids (MPI CPfS), Dresden 01187, Germany}

\author{Lin Jiao}
\email[Corresponding author: ]{lin.jiao@zju.edu.cn}
\affiliation  {New Cornerstone Science Laboratory, Center for Correlated Matter
	and School of Physics, Zhejiang University, Hangzhou 310058, China}

\author{Michael Smidman}
\email[Corresponding author: ]{msmidman@zju.edu.cn}
\affiliation  {New Cornerstone Science Laboratory, Center for Correlated Matter and School of Physics, Zhejiang University, Hangzhou 310058, China}

\author{Huiqiu Yuan}
\email[Corresponding author: ]{hqyuan@zju.edu.cn}
\affiliation  {New Cornerstone Science Laboratory, Center for Correlated Matter and School of Physics, Zhejiang University, Hangzhou 310058, China}
\affiliation  {Institute of Fundamental and Transdisciplinary Research, Zhejiang University, Hangzhou 310058, China}
\affiliation  {State Key Laboratory of Silicon and Advanced Semiconductor Materials, Zhejiang University, Hangzhou 310058, China}


\begin{abstract}

Unconventional superconductivity typically occurs near magnetic instabilities, and the corresponding spin fluctuations are widely believed to play a crucial role in mediating electron pairing. UTe$_2$ is a promising candidate for exhibiting multiple spin-triplet superconducting phases when tuning with applied pressure and magnetic fields, but the nature of the magnetism driving these unconventional pairing states is undetermined. Our measurements of UTe$_2$ under applied pressures and magnetic fields reveal the presence of a magnetic order hidden within the pressure-induced superconducting dome, which vanishes together with the superconductivity once there is sufficiently high pressure to induce the three-dimensional antiferromagnetic phase. Extrapolation of the phase boundary of the hidden magnetic order, which is most likely antiferromagnetic in nature, points to a zero-temperature quantum critical point that coincides with the maximum transition temperature of the pressure-induced superconducting dome, suggesting that it could corresponds to the parent magnetic phase of the critical antiferromagnetic spin fluctuations driving the triplet superconductivity. 
These findings advance the understanding of the interplay of magnetism and superconductivity in an exemplar candidate triplet superconductor, which is necessary for revealing the microscopic origin of the different unconventional superconducting phases.

\end{abstract}

\maketitle

Spin-triplet superconductors, where the spins of the Cooper pairs form a symmetric $S$=1 configuration, are a much sought after type  of unconventional superconductor, partly because they may host topological superconductivity with applications in quantum information \cite{satoTopologicalSuperconductorsReview2017}. A handful of heavy fermion candidates
 for odd-parity superconductivity have been identified,  including UGe$_2$ \cite{saxenaSuperconductivityBorderItinerantelectron2000}, URhGe \cite{aokiCoexistenceSuperconductivityFerromagnetism2001}, UCoGe \cite{huySuperconductivityBorderWeak2007}, UPt$_3$ \cite{aeppliMagneticOrderDifferent1989,joyntSuperconductingPhasesUPt2002}, UNi$_2$Al$_3$ \cite{ishidaSpintripletSuperconductivity$mathrmUNi_2mathrmal_3$2002} and CeSb$_2$ \cite{squireSuperconductivityConventionalPauli2023c,shanEmergentFerromagneticLadder2025c}, posing the question as to the nature of the spin fluctuations favorable for triplet pairing. To this end it has been crucial to characterize the nature of magnetically ordered phases coexistent or in close proximity to the superconductivity, where in  UGe$_2$, URhGe, and UCoGe there is coexistence with ferromagnetic (FM) order \cite{saxenaSuperconductivityBorderItinerantelectron2000,aokiCoexistenceSuperconductivityFerromagnetism2001,huySuperconductivityBorderWeak2007}, while UPt$_3$ exhibits a weak antiferromagnetic (AFM) state \cite{aeppliMagneticOrderDifferent1989}.


UTe$_2$ is another promising candidate for being a spin-triplet superconductor, exhibiting a critical temperature $T_{\text{sc}} = \SI{2.1}{K}$ and a large anisotropic upper critical field $H_{\text{c2}}$ that exceeds the Pauli limit along all crystallographc directions \cite{ranNearlyFerromagneticSpintriplet2019a,sakaiSingleCrystalGrowth2022}. When a magnetic field is applied along the $b$-axis, the superconducting state evolves from a low-field phase denoted SC1 into a distinct high-field phase (SC2) \cite{ranExtremeMagneticFieldboosted2019,wuEnhancedTripletSuperconductivity2024,knebelFieldreentrantSuperconductivityClose2019,rosuelFieldinducedTuningPairing2023}. This SC2 phase abruptly disappears once the system enters the field-polarized state \cite{ranExtremeMagneticFieldboosted2019,wuEnhancedTripletSuperconductivity2024,knebelFieldreentrantSuperconductivityClose2019,rosuelFieldinducedTuningPairing2023}, but for specific field orientations, reentrant superconductivity emerges \cite{ranExtremeMagneticFieldboosted2019,wuEnhancedTripletSuperconductivity2024}, forming a halo in the phase diagram \cite{lewinHighfieldSuperconductingHalo2025}.
Spin-triplet pairing in UTe$_2$ is strongly suggested by several NMR experiments \cite{matsumuraLargeReductionAaxis2023,kinjoSuperconductingSpinReorientation2023,nakamineAnisotropicResponseSpin2021,nakamineSuperconductingPropertiesHeavy2019}, while there have been various proposals for the specific order parameter symmetry \cite{,liObservationOddparitySuperconductivity2025,hayesRobustNodalBehavior2025,wangOddparityQuasiparticleInterference2025,FullyGappedPairinga,theussSinglecomponentSuperconductivityUTe22024,jiaoChiralSuperconductivityHeavyfermion2020,azariAbsenceSpontaneousMagnetic2023a,guPairWaveFunction2025,ajeeshFateTimereversalSymmetry2023a}.
Importantly, the presence of both AFM \cite{duanIncommensurateSpinFluctuations2020,knafoLowdimensionalAntiferromagneticFluctuations2021} and FM \cite{tokunagaLongitudinalSpinFluctuations2023,sundarCoexistenceFerromagneticFluctuations2019} spin fluctuations at ambient pressure poses the question as to which are critical for driving the superconducting pairing.



Upon applying pressure,  an additional superconducting phase emerges above \SI{0.2}{GPa} situated at temperatures above SC1 \cite{thomasEvidencePressureinducedAntiferromagnetic2020,braithwaiteMultipleSuperconductingPhases2019,aokiMultipleSuperconductingPhases2020,wuMagneticSignaturesPressureInduced2025,vasinaQuantitativeThermodynamicStudy2026}, which is the same as the ambient-pressure high-field SC2 phase \cite{vasinaConnectingHighFieldHighPressure2025}. NMR under pressure shows that the $b$-axis Knight shift remains almost unchanged within the SC2 phase, indicating that this is a spin-triplet state distinct from SC1 with a different orientation of the superconducting order parameter \cite{kinjoSuperconductingSpinReorientation2023}. While the transition into SC1 is monotonically suppressed by pressure, the SC2 phase exhibits a dome-shape, and superconductivity abruptly disappears above $P_{\text{c1}} = \SI{1.5}{GPa}$, where there is the appearance of magnetic order \cite{thomasEvidencePressureinducedAntiferromagnetic2020,braithwaiteMultipleSuperconductingPhases2019,aokiMultipleSuperconductingPhases2020,wuMagneticSignaturesPressureInduced2025,vasinaQuantitativeThermodynamicStudy2026}. Neutron diffraction reveals that this magnetism corresponds to three-dimensional (3D) long-range AFM order \cite{knafoIncommensurateAntiferromagnetismUTe22025}, with a different wave vector to the ambient pressure AFM spin fluctuations \cite{duanIncommensurateSpinFluctuations2020,knafoLowdimensionalAntiferromagneticFluctuations2021}. Another transition is detected above the 3D-AFM phase, designated as weak magnetic order (WMO) \cite{aokiFieldinducedSuperconductivitySuperconducting2021a,liMagneticPropertiesPressure2021}, which may correspond to short range ordering.

In our work, we reveal a new magnetic phase under pressure in UTe$_2$ which is hidden in the superconducting dome. The magnetic-field response of this hidden magnetic order (HMO) is in line with antiferromagnetism and, most likely, is connected to the low dimensional AFM spin fluctuations present at ambient and applied pressures. On the other hand, applying a magnetic field just below $P_{\text{c1}}$ induces the WMO and 3D-AFM transitions while destroying superconductivity, indicating competition of these magnetic phases with superconductivity. As will become clear below, these findings motivate a new scenario for the interplay of superconductivity and magnetism in UTe$_2$, in which the spin-triplet SC2 phase is driven by a hidden AFM instability associated with a quantum critical point.





\begin{figure}
  \centering
  \includegraphics[angle=0,width=0.45 \textwidth]{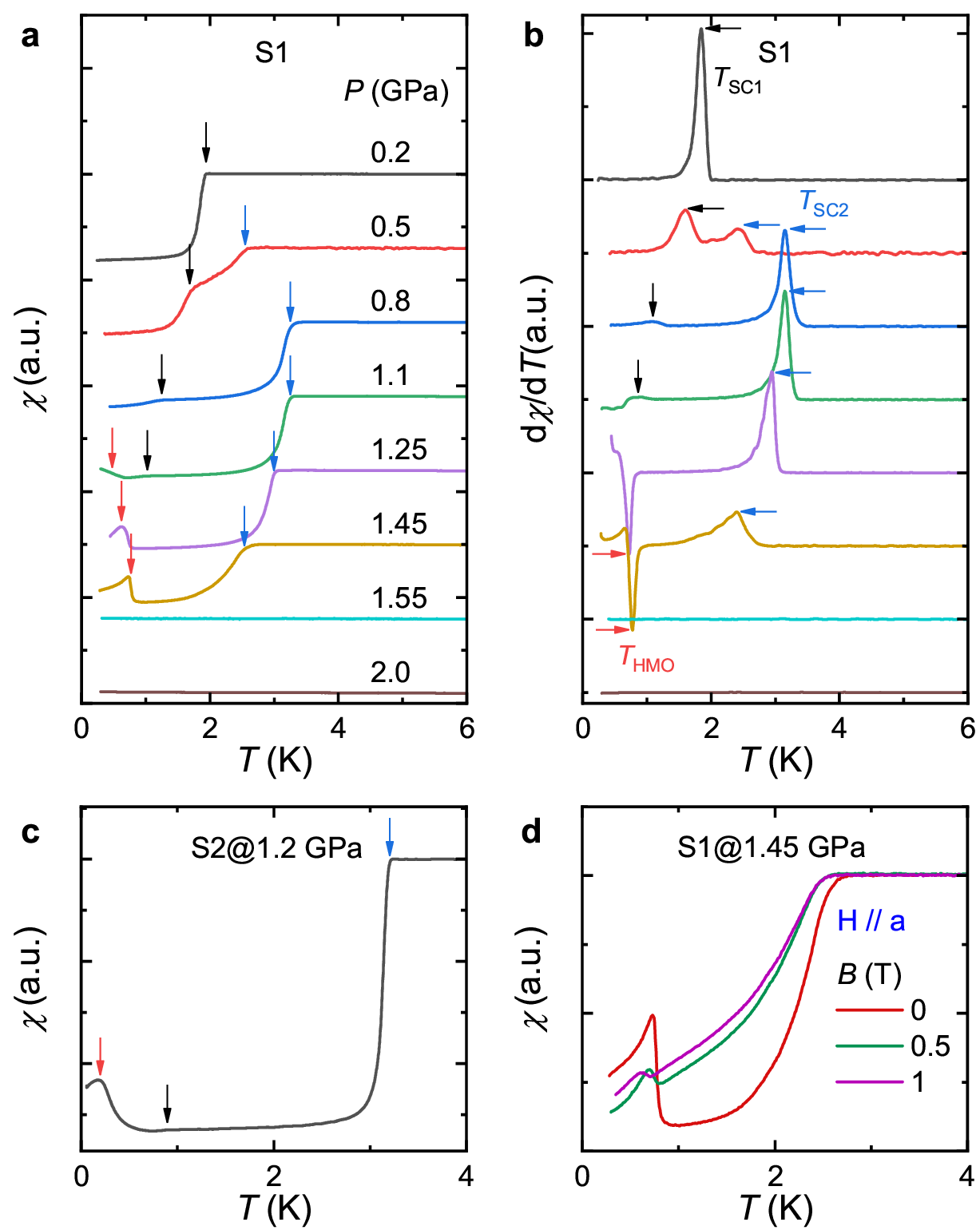}
	\vspace{1mm}
  \caption{ \textbf{Magnetic susceptibility of UTe$_2$  under pressure showing hidden magnetic order.} Temperature dependence of the \textbf{a,} ac magnetic susceptibility $\chi (T)$, and \textbf{b,} the corresponding derivatives $d\chi/dT(T)$, of UTe$_2$ sample S1 under various pressures. The $\chi (T)$ curves are shifted vertically for clarity. The black, blue and red arrows indicate the transitions to the SC1, SC2 and HMO phases, respectively, which are defined by the peak or minima  positions in the derivative.  $\chi (T)$ of  \textbf{c,} sample S2 at \SI{1.2}{GPa}, and \textbf{d,} S1 under various magnetic fields at \SI{1.45}{GPa}.  
  }
 
  \label{fig1}
	\vspace{-10pt}
\end{figure}

Figure \ref{fig1}\textbf{a} displays the temperature dependence of the ac magnetic susceptibility $\chi (T)$ under various pressures for sample S1. At \SI{0.2}{GPa}, only a single sharp drop is observed at \SI{1.85}{K}, corresponding to the superconducting transition into SC1 (black arrow). At \SI{0.5}{GPa}, two superconducting transitions are observed consistent with previous reports \cite{thomasEvidencePressureinducedAntiferromagnetic2020,braithwaiteMultipleSuperconductingPhases2019,aokiMultipleSuperconductingPhases2020,wuMagneticSignaturesPressureInduced2025,vasinaQuantitativeThermodynamicStudy2026}. Here the lower transition to the SC1 phase decreases with pressure, and cannot be discerned above \SI{1.1}{GPa}, while the higher transition to the SC2 phase (blue arrows) increases from \SI{2.42}{K} at \SI{0.5}{GPa} to a maximum of \SI{3.15}{K} at \SI{1.1}{GPa}, and then  decreases to \SI{2.40}{K} at \SI{1.45}{GPa}, corresponding to the reported dome in the phase diagram \cite{thomasEvidencePressureinducedAntiferromagnetic2020,braithwaiteMultipleSuperconductingPhases2019,aokiMultipleSuperconductingPhases2020,wuMagneticSignaturesPressureInduced2025,vasinaQuantitativeThermodynamicStudy2026}. 

Importantly, at 1.1 GPa an additional low temperature anomaly emerges marked by the red arrows, which does not show the diamagnetic shift of the superconducting transitions, but has a pronounced peak characteristic of a magnetic phase (denoted HMO) within the superconducting state. This feature, which is reproducible in another sample S2 (Fig.~\ref{fig1}\textbf{c}), shifts to higher temperature with increasing pressure, the opposite pressure evolution to the SC2 transition in this pressure range. Meanwhile at 1.55 GPa, both these magnetic and superconducting transitions disappear, and  $\chi (T)$ is nearly temperature-independent, which is consistent with previous reports where no signatures of the 3D-AFM or WMO phases are detected in the ac magnetic susceptibility  \cite{wuMagneticSignaturesPressureInduced2025}. Moreover, upon applying magnetic fields, the  HMO transition shifts to lower temperatures (Fig.~\ref{fig1}\textbf{d}), as expected for an AFM phase.

\begin{figure}
	\includegraphics[angle=0,width=0.48\textwidth]{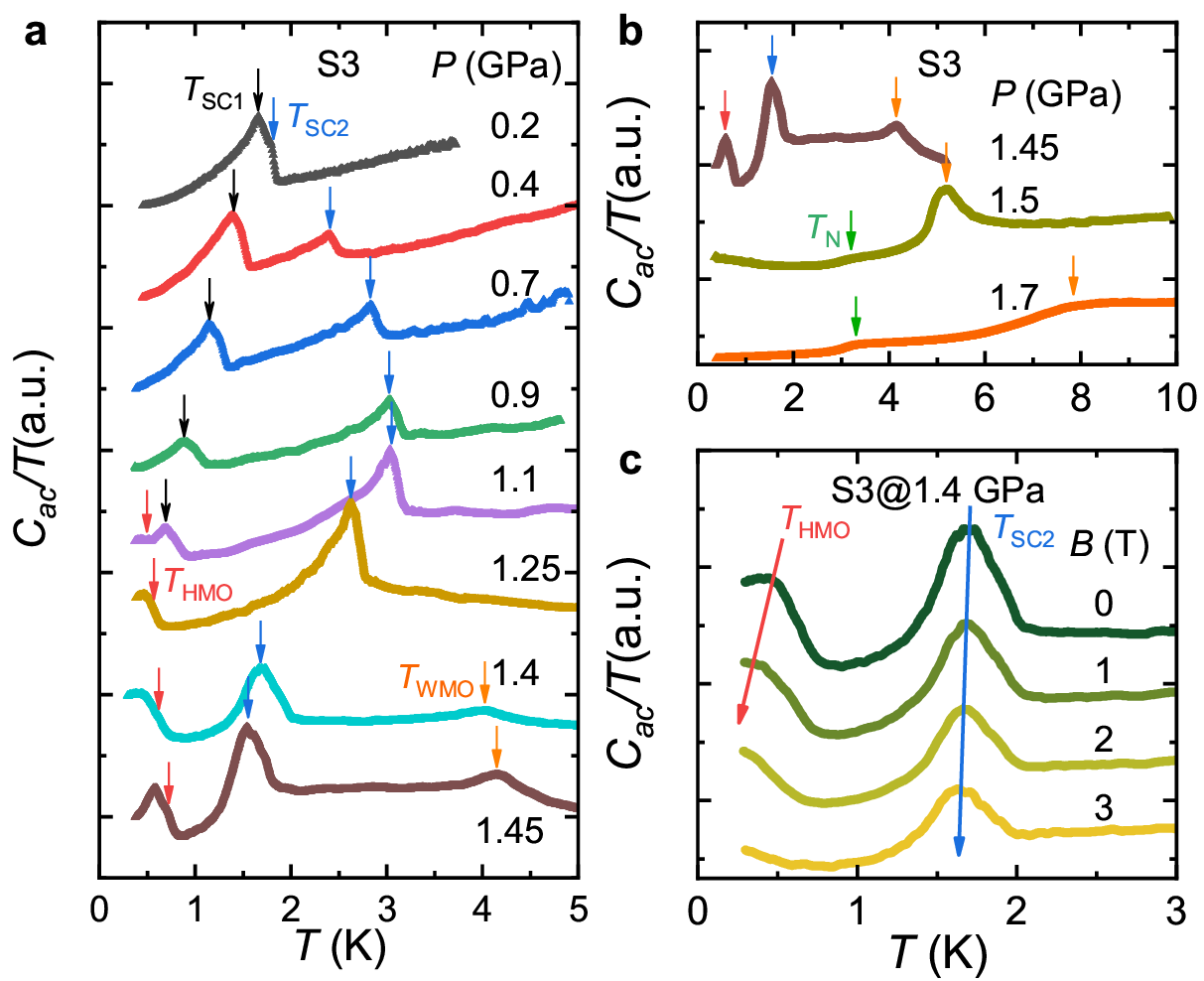}
	\caption{ \textbf{AC heat capacity of UTe$_2$  under pressure.} Temperature dependence of the ac heat capacity as $C/T$ for sample S3 under pressures \textbf{a,} up to 1.45 GPa, and \textbf{b,} above 1.45 GPa. The transitions to the SC1 (black), SC2 (blue), HMO (red), 3D-AFM (green) and WMO (orange) phases are marked by arrows.  \textbf{c,} $C/T$ versus temperature at \SI{1.4}{GPa} under various applied magnetic fields, where the trend of the HMO and SC2 transitions are shown by arrows.
	}
	\label{fig2}
	\vspace{-10pt}
\end{figure}

\begin{figure}
	\includegraphics[angle=0,width=0.45\textwidth]{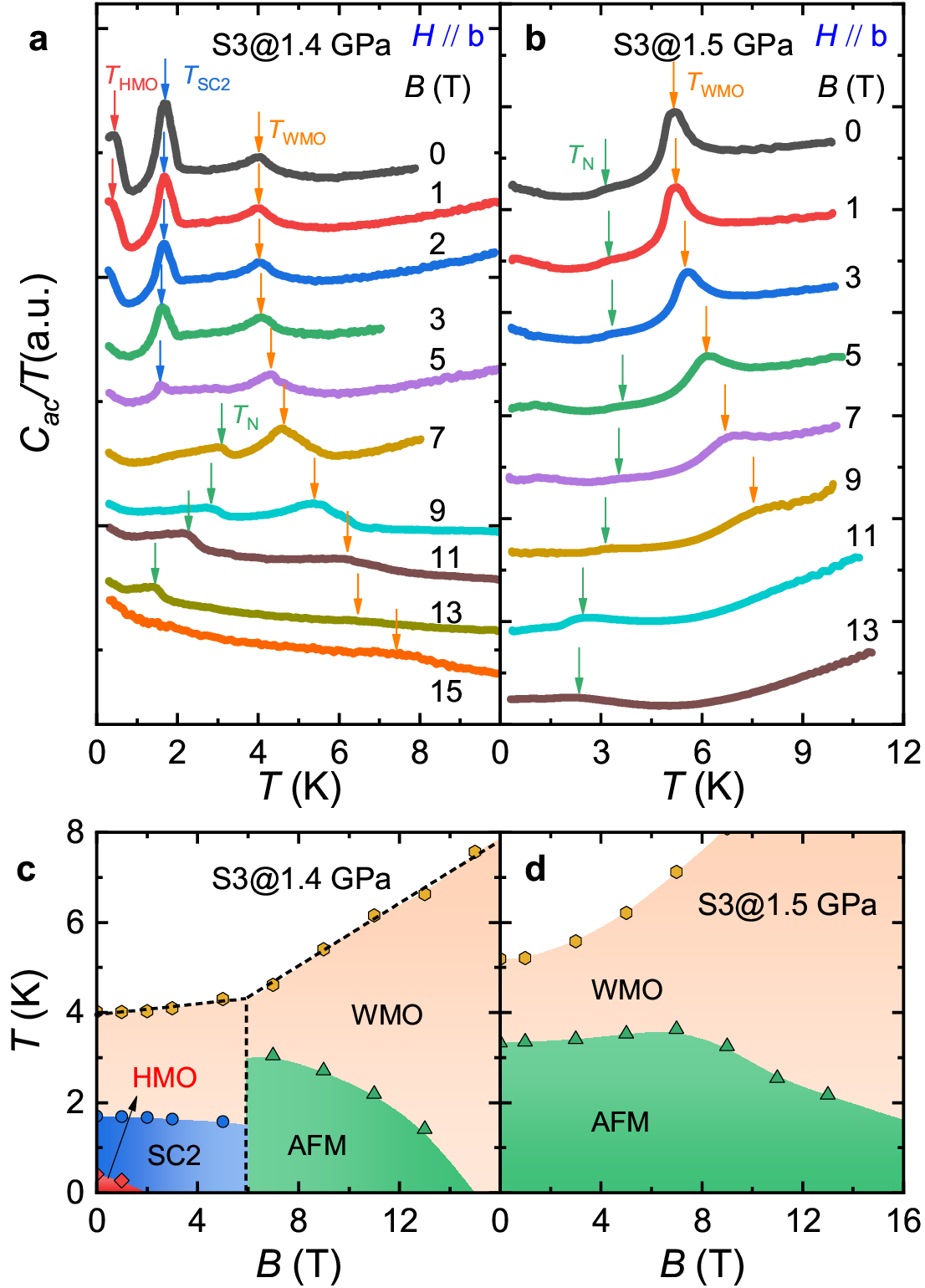}
	\caption{\textbf{Competition between AFM/WMO and SC2 under pressure revealed by applied magnetic fields.} 
	Temperature dependence of $C/T$  of S3 under various magnetic fields applied along the $b$-axis at \textbf{a,} \SI{1.4}{GPa}, and \textbf{b,} \SI{1.5}{GPa}.
	Temperature--magnetic field phase diagrams at \textbf{c,}  \SI{1.4}{GPa} and \textbf{d,}  \SI{1.5}{GPa}. The dashed lines are guides to the eye. Applying a magnetic field just below $P_{\text{c1}}$ leads to the abrupt disappearance of superconductivity, which is replaced by the 3D-AFM phase.
	}
	\label{fig3}
	\vspace{-10pt}
\end{figure}

The existence of an additional magnetic transition hidden in the superconducting phase is further confirmed by ac calorimetry measurements under pressure shown in  Fig.~\ref{fig2}. Here, the two superconducting transitions can already be resolved at \SI{0.2}{GPa}, which further split at higher pressures, where the lower transition to the SC1 phases decreases with pressure, and the higher transition to SC2 increases to a maximum around 1.1 GPa before decreasing, consistent with the ac susceptibility results. At \SI{1.1}{GPa}, an anomaly develops below the lower superconducting transition at $\SI{0.55}{K}$, coinciding with the feature of the HMO transition found in the ac susceptibility. Upon increasing the pressure the anomaly shifts to higher temperature and evolves into a pronounced peak, which shifts to lower temperatures with increasing magnetic fields up to 3 T [Fig.~\ref{fig2}\textbf{c}]. As shown in Fig.~\ref{fig2}\textbf{b}, both the superconducting and HMO transitions disappear at $P_{\text{c1}} = \SI{1.5}{GPa}$, but anomalies associated with the WMO and 3D-AFM (labelled $T_{\text{N}}$) transitions are now observed. The WMO transition actually begins to be detectable at a lower pressure of 1.4 GPa and shifts to higher temperature with increasing pressure, while $T_{\text{N}}$ only increases slightly from \SI{3.2}{K} at \SI{1.5}{GPa} to \SI{3.3}{K} at \SI{1.7}{GPa}, consistent with previous reports \cite{thomasEvidencePressureinducedAntiferromagnetic2020,braithwaiteMultipleSuperconductingPhases2019,aokiMultipleSuperconductingPhases2020,vasinaQuantitativeThermodynamicStudy2026}. Meanwhile the change in the magnitude and shape of the specific heat
anomaly just below $P_{\text{c1}}$ in Ref.~\cite{vasinaQuantitativeThermodynamicStudy2026} is also consistent with our findings of a change of ground state. Consequently, both ac magnetic susceptibility and ac calorimetry consistently show a likely AFM order existing hidden in the superconducting phase, that disappears with the superconductivity once 3D-AFM order onsets at $P_{\text{c1}}$.

To further understand the interplay of these phases, results of ac calorimetry measurements in $b$-axis magnetic fields up to 15 T are displayed in Fig.~\ref{fig3} under pressures of \SI{1.4}{GPa}, just below $P_{\text{c1}}$, and at $P_{\text{c1}}=1.5$~GPa.  At \SI{1.4}{GPa}, the HMO, SC2 and WMO transitions are all observed in zero-field, where the former two are suppressed with increasing field, with the peak of the HMO transition being below the lowest measured temperature at 2 T,   while the WMO transition shifts slightly to higher temperatures. At 7 T however, the SC2 transition abruptly disappears, and a pronounced peak corresponding to the $T_{\text{N}}$ of the 3D-AFM  transition emerges. In higher fields,  $T_{\text{N}}$ also decreases, whereas the transition to the WMO phase shows a more rapid increase with field once SC2 vanishes, reaching \SI{7.56}{K} at \SI{15}{T}, as shown by the phase diagram in Fig.~\ref{fig3}\textbf{c}. At \SI{1.5}{GPa}, where the 3D-AFM state is present at zero-field, $T_{\text{N}}$ shows a non-monotonic field dependence, with a weak increase up to  \SI{7}{T}, followed by a decrease at higher fields, reaching \SI{2.2}{K} at \SI{13}{T}, as shown in Fig.~\ref{fig3}\textbf{d}. In contrast, the WMO transition increases more rapidly at low fields as compared to the results at \SI{1.4}{GPa}, and becomes too broad to be detected above \SI{9}{T}, in line with Ref.~\cite{knebelAxisElectricalTransport2024}.

\begin{figure}
	\centering
	\includegraphics[angle=0,width=0.48 \textwidth]{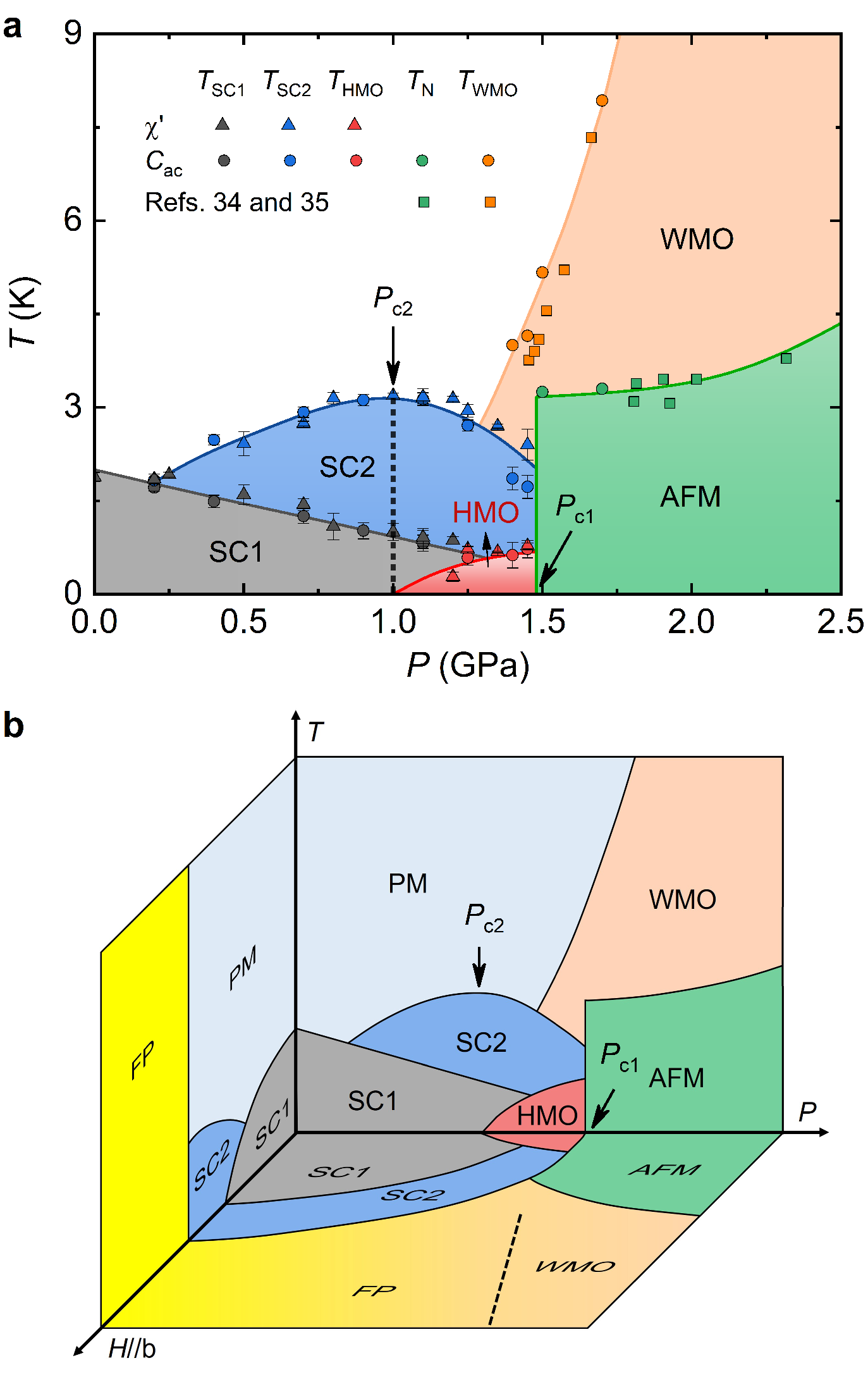}
	\caption{\textbf{Phase diagram of UTe$_2$.} 
		\textbf{a} $T$--$P$ phase diagram of UTe$_2$ up to 2.5 GPa. The triangular and circular symbols represent the data from ac magnetic susceptibility and ac calorimetry measurements respectively, while square symbols denote the transitions to the 3D-AFM and WMO phase extracted from Refs.~\onlinecite{thomasEvidencePressureinducedAntiferromagnetic2020,braithwaiteMultipleSuperconductingPhases2019}.
		\textbf{b} Schematic 3D phase diagram of UTe$_2$. PM denotes the paramagnetic phase, and FP denotes the field-polarized phase.
	}
	\label{fig4}
	\vspace{-10pt}
\end{figure}

The revised temperature--pressure phase diagram incorporating the hidden magnetic order within the SC2 phase is displayed in Fig.~\ref{fig4}\textbf{a}. 
The critical pressure ($P_{\text{c2}} \approx \SI{1.0}{GPa}$) at which the HMO emerges coincides with the maximum of the SC2 dome and occurs before SC1 is fully suppressed by pressure. 
The HMO transition increases with pressure, and the WMO phase emerges slightly below $P_{\text{c1}}$, while above $P_{\text{c1}}$, both SC2 and HMO are replaced by the 3D-AFM phase.

Notably, both FM and AFM spin fluctuations have been reported at ambient \cite{duanIncommensurateSpinFluctuations2020,knafoLowdimensionalAntiferromagneticFluctuations2021,tokunagaLongitudinalSpinFluctuations2023,sundarCoexistenceFerromagneticFluctuations2019} and applied pressure \cite{ambikaPossibleCoexistenceAntiferromagnetic2022,vijayanambikaEnhancementAntiferromagneticSpin2026,kinjoDrasticChangeMagnetic2022}, where the former have been associated with the magnetic coupling within the U-ladders, while the latter have been ascribed to inter-ladder coupling \cite{ambikaPossibleCoexistenceAntiferromagnetic2022,knafoLowdimensionalAntiferromagneticFluctuations2021}. However, the incommensurate wave vector characterizing the low-dimensional AFM fluctuations is along \textbf{b*} \cite{knafoLowdimensionalAntiferromagneticFluctuations2021}, which is different from that of the 3D-AFM order appearing above \SI{1.5}{GPa} \cite{knafoIncommensurateAntiferromagnetismUTe22025}. 
Our findings therefore point to the AFM spin fluctuations present at ambient and low pressures being associated with the HMO phase instead.
Since applied pressure has been found to suppress FM fluctuations, and enhance the aforementioned AFM fluctuations \cite{vijayanambikaEnhancementAntiferromagneticSpin2026}, which mirrors the respective pressure dependences of SC1 and SC2, it has been suggested that SC1 and SC2 correspond to different spin triplet pairing states driven by FM and AFM fluctuations, respectively \cite{vijayanambikaEnhancementAntiferromagneticSpin2026,teiPairingSymmetriesMultiple2024}. This picture is supported by calculations based on a periodic Anderson model, which predict that AFM fluctuations can drive triplet superconductivity in UTe$_2$ when the wave vector is orientated along \textbf{b*} \cite{hakunoMagnetismSuperconductivityMixeddimensional2024}, whereas it is not supported for the wave vector of the 3D-AFM order \cite{knafoIncommensurateAntiferromagnetismUTe22025}. Moreover, the HMO phase boundary extrapolates to zero temperature at $P_{\text{c2}} \approx \SI{1.0}{GPa}$, which coincides with the maximum transition temperature of the SC2 phase, as indicated by dashed line in Fig.~\ref{fig4}\textbf{a}. These considerations therefore suggest that SC2 may be driven by critical AFM fluctuations originating in a quantum critical point (QCP) of the HMO phase.  Indeed, signatures of quantum criticality have been observed slightly above $P_{\text{c2}}$ but below $P_{\text{c1}}$ \cite{braithwaiteMultipleSuperconductingPhases2019,thomasEvidencePressureinducedAntiferromagnetic2020,vasinaQuantitativeThermodynamicStudy2026}. 
Specifically, the $A$ coefficient obtained from the fit $\rho = \rho_0 + A T^2$ diverges near $P_{\text{c2}}$ \cite{braithwaiteMultipleSuperconductingPhases2019}. In the same pressure region, the Sommerfeld coefficient $\gamma$ reaches a maximum \cite{vasinaQuantitativeThermodynamicStudy2026} and the resistivity exhibits a linear temperature dependence \cite{thomasEvidencePressureinducedAntiferromagnetic2020}. This ``strange-metal'' behavior may indicate a Mott type AFM QCP \cite{huQuantumCriticalMetals2024}, but this, as well as whether Mott-like AFM spin fluctuations drive the spin-triplet pairing of SC2, remain to be clarified by future microscopic studies.

Previously a QCP around $P_{\text{c2}}$ was instead ascribed to being related to the WMO phase \cite{thomasEvidencePressureinducedAntiferromagnetic2020}. However, our field-dependent measurements at 1.4 GPa, just below $P_{\text{c1}}$, not only reveal competition between the 3D-AFM and SC2 phases [Fig.~\ref{fig3}\textbf{c}], but also competition between the WMO and SC2 phases. As indicated by the dashed lines in Fig.~\ref{fig3}\textbf{c}, once the field has suppressed superconductivity, the temperature of the WMO transition shows a more rapid increase with field.
Our findings that applying $b$-axis fields just below $P_{\text{c1}}$ leads to the sudden replacement of SC2 with the 3D-AFM phase can be compared to the results at ambient and lower pressures for this field direction illustrated schematically in Fig.~\ref{fig4}\textbf{b}, where SC2 abruptly disappears at a metamagnetic transition $H_{\text{m}}$ to the field-polarized state \cite{ranExtremeMagneticFieldboosted2019,wuEnhancedTripletSuperconductivity2024,knebelFieldreentrantSuperconductivityClose2019,rosuelFieldinducedTuningPairing2023}. This metamagnetic transition, and hence the critical field for superconductivity, are drastically suppressed by pressure \cite{knebelAnisotropyUpperCritical2020,vasinaConnectingHighFieldHighPressure2025,knebelAxisElectricalTransport2024}, coinciding with the $b$-axis changing from the hard to easy axis by \SI{1.7}{GPa} \cite{liMagneticPropertiesPressure2021}. In contrast, applying magnetic fields along the hard $c$-axis just above $P_{\text{c1}}$ has the opposite effect, whereby 3D-AFM order is replaced by reentrant superconductivity \cite{aokiFieldinducedSuperconductivitySuperconducting2021a}. Therefore, together these observations indicate that enhanced spin polarization along the $b$-axis strongly disfavors the SC2 phase.

Consequently, these findings can provide a significantly altered perspective on the interplay of the multiple magnetic and superconducting phases of UTe$_2$, identifying a hidden antiferromagnetism within the superconducting dome as being a strong candidate for the progenitor of the spin-triplet SC2 phase. Detailed characterizations of the spin fluctuations related to the HMO, as well as the other magnetic phases, will be crucial for resolving their impact on the Cooper-pair formation in the different phases of superconductivity in UTe$_2$.

\section{Acknowledgments}

We acknowledge fruitful discussions with P. Coleman, C. Cao and L.H. Hu. This work was supported by the National Key R\&D Program of China (No. 2022YFA1402200, No. 2023YFA1406303 and No. 2023YFA1406103), the National Science Foundation of China (No. W2511006, No. 12494592, No. 12550401, No. 12374151 and No. U2430209), the Zhejiang Provincial Natural Science Foundation of China (No. LRG26A040001 and No. LR25A040003), and the New Cornerstone Science Foundation (No. NCI202509).

\section{Methods}

\subsection{Sample growth and preparation}

High-quality UTe$_2$ single crystals were synthesized via a molten salt flux (MSF) technique employing an equimolar NaCl-KCl mixture (99.99\% purity, Alfa Aesar) as a flux \cite{sakaiSingleCrystalGrowth2022}. Uranium metal pieces were initially etched in nitric acid to eliminate surface oxides. Under an argon atmosphere, tellurium pieces (Te, 99.999\% purity, Alfa Aesar) were combined with uranium in a U:Te molar ratio of 1:1.65, along with NaCl-KCl flux with a U:salt molar ratio of 1:60. The reactants were loaded into a carbon crucible lined with quartz wool to prevent material loss during heating. The crucible was placed in a quartz tube and heated to $180^\circ\text{C}$ under high vacuum (\textless \SI{5e-4}{Pa}) for dehydration. Subsequently, the ampoule was vacuum-sealed and placed in a box furnace, where the mixture was heated to $950^\circ\text{C}$ over 24 hours and maintained at this temperature for 48 hours. It was then cooled gradually at a rate of $0.03^\circ\text{C}$/min to $650^\circ\text{C}$, held at this temperature for 48 hours, and finally allowed to cool naturally to ambient temperature. 

\subsection{Ac susceptibility and ac calorimetry measurements}

Ac susceptibility measurements were performed using a three-coil setup consisting of nested primary (drive), secondary (pickup), and compensation coils.
$\chi (T)$ was obtained from the real part of a lock-in measurement of the voltage induced across the pickup coil. Measurements were conducted with an ac excitation frequency of \SI{1333}{Hz} and current of \SI{200}{\micro\ampere}.

For ac calorimetry measurements, a \SI{20}{\ohm} constantan heater is used to generate a small temperature oscillation and a chromel-AuFe (0.07\%) thermocouple is utilized to pick up the heat capacity signal. To achieve a better signal-to-noise ratio, an excitation current of \SI{1.4}{mA} is used above \SI{2}{K}, while a current of \SI{0.9}{mA} is employed for measurements below \SI{2}{K} to avoid heating effects. 

Measurements under pressure were performed utilizing a \SI{4}{mm} bore piston-cylinder cell, with Daphne oil 7373 as the pressure-transmitting medium. The pressure was calibrated \textit{in situ} at low temperature by monitoring the superconducting transition of a lead manometer. Low temperature measurements were performed in Teslatron-PT system with an Oxford $^3\text{He}$ refrigerator and an Oxford $^3\text{He}$/dilution cryostat.

\section{Author contributions}

 H.Y. conceived the experiments. D.Y., B.Z., C.Z., Q.C. and S.T. synthesized the single crystals. K.Y., L.W., Y.Z. and Y.C. performed the ac magnetic susceptibility and ac calorimetry measurements under pressure. K.Y., L.J., M.S. and H.Y. analyzed the data. K.Y., Y.L., X.L. S.F., L.J., M.S. and H.Y. wrote the paper with input from all authors.  

\section{Competing interests}

The authors declare no competing interests.

\section{Additional information}

\textbf{Correspondence and requests for materials} should be addressed to Lin Jiao, Michael Smidman or Huiqiu Yuan.

\bibliography{UTe2_HAFO}

\end{document}